\documentclass{osa-article}
\usepackage{cite}
\usepackage{amsmath}
\usepackage{braket}
\usepackage[]{graphicx}
\usepackage{bm}
\usepackage{booktabs}
\usepackage{makecell}
\usepackage{multirow}
\usepackage{color}
\usepackage[pagewise]{lineno}
\newlength{\figwidth} 
\newcommand{\beq}{\begin{equation}}
\usepackage{comment}
\usepackage{verbatim}
\usepackage{url}

\newcommand{\beql}[1]{\begin{equation}\label{#1}}
\newcommand{\eeq}{\end{equation}}
\newcommand{\bsp}{\begin{split}}
\newcommand{\esp}{\end{split}}
\newcommand{\Eq}[1]{Eq.~(\ref{#1})}

\journal{oe}
\articletype{Research Article}

\begin{document}

\title{Gauge-field description of Sagnac frequency shift and mode hybridization in a rotating cavity}

\author{Hongkang Shi,\authormark{1} Zhongfei Xiong,\authormark{1} Weijin Chen,\authormark{1} Jing Xu,\authormark{1,3} Shubo Wang,\authormark{2,4} and Yuntian Chen,\authormark{1,3,5}}

\address{\authormark{1}School of Optical and Electronic Information, Huazhong University of Science and Technology, Wuhan, China\\
\authormark{2}Department of Physics, City University of Hong Kong, Hong Kong, China\\
\authormark{3}Wuhan National Laboratory for Optoelectronics, Huazhong University of Science and Technology,
Wuhan, China\\}
\email{\authormark{4}shubwang@cityu.edu.hk} 
\email{\authormark{5}yuntian@hust.edu.cn} 

\begin{abstract}
Active optical systems can give rise to intriguing phenomena and applications that are not available in conventional passive systems. Structural rotation has been widely employed to achieve non-reciprocity or time-reversal symmetry breaking. Here, we examine the quasi-normal modes and scattering properties of a two-dimensional cylindrical cavity under rotation. In addition to the familiar phenomenon of Sagnac frequency shift, we observe the the hybridization of the clockwise(CW) and counter-clockwise(CCW) chiral modes of the cavity controlled by the rotation. The rotation tends to suppress one chiral mode while amplify the other, and it leads to the variation of the far field. The phenomenon can be understood as the result of a synthetic gauge field induced by the rotation of the cylinder. We explicitly derived this gauge field and the resulting Sagnac frequency shift. The analytical results are corroborated by finite element simulations. Our results can be applied in the measurement of rotating devices by probing the far field.
\end{abstract}
\section{Introduction}

 The rotating cavity is a rich subject with far-reaching implications and has many applications such as optical gyroscopes\cite{sarma2012wavelength,sunada2007design,mignot2009single}, as evidenced by the long record of literature in the past four decades. The rotation induced effects, such as Sagnac frequency splitting and the modified far-field pattern, can be interpreted in various ways by the stationary observer (inertial frame) and the moving observer (reference frame rotating with the cavity). Coincidentally, it is found that the first order approximation\cite{post1967sagnac} in either inertial frame or rotating frame yields consistent results on the Sagnac frequency shift by using the proper non-relativistic constitutive relation for the moving medium. Interestingly, Cook and co-workers compare the time-independent Schr\"ondinger equation with the wave-equation for the electric field in moving medium up to the first order\cite{cook1995fizeau}, and concluded that the velocity of the moving medium $\bm v$ and the vorticity $\nabla \times \bm v$ play the roles of the vector potential and the effective magnetic field that bend light, analogous to electrons travelling in magnetic field. The similar concept as well as the first-order treatment are also used to examine the Aharonov-Bohm effect of light associated with a rotating rod immersed in a viscous fluid\cite{vieira2014aharonov}. Considering the recent development of topological photonics, it is theoretically appealing to have a gauge field description\cite{bliokh2004modified,fang2012photonic,liu2015gauge,chen2018non,wang2018topological}of the optical properties of rotating cavities beyond the first order approximation.

 In this paper, with Minkowski's postulation of the modified constitutive relation of moving  medium\cite{minkowski1908h}, we revisit the rotating rod hosted in vacuum and develop a physically transparent yet rigorous description for the Sagnac frequency splitting and modal hybridization of the cavity by constructing the effective gauge field induced by rotation. In particular, we consider finite quality(Q)-factor cavity modes supported by high refractive index dielectric rod. The rotating cavity can be equivalently treated as stationary cavity with bianisotropic response, which can be further solved by analytical or numerical approaches. From the first principle of Maxwell's equation, we analytically derive the rotation-induced gauge field, the effective magnetic field, and the Sagnac frequency splitting $\Delta \omega$, which are further benchmarked against full-wave finite-element simulations. At low rotation speed, as $\Delta \omega$ is small, the spectra of the CW and CCW resonances overlap, leading to rotation-induced evolution of far-field emission patterns as explained qualitatively by Ge et al\cite{ge2015rotation}. The spectra overlapping is essentially due to the modal hybridization between the CW and CCW modes, which is semi-analytically described by the coherent superposition of the two quasi-normal modes in this paper.

 The paper is organized as follows. In Section 2, we give the detailed derivations of rotation-induced gauge field and effective magnetic field as well as frequency splitting $\Delta \omega$.  The simulation results and discussions are provided in Section 3. Finally, the paper is concluded in Section 4.

\section{Fundamentals of rotation-induced gauge field}

\subsection{  Synthetic gauge field and effective magnetic field in a rotating cavity}

Figure 1(a) shows a two-dimensional (2D) view of a cylinder rotating around the $z$  axis with angular velocity $\Omega$, where $(r,\theta)$ are, respectively, the radial and azimuthal coordinates, and $\hat {\bm e}_r$ and $\hat {\bm e}_\theta$ are the corresponding unit vectors. The linear velocity of the rotating cylinder is $\bm v = \Omega r \hat {\bm e}_\theta$ at a radius $r$. The modified constitutive relation \cite{minkowski1908h} of the moving medium in the stationary frame is given by, 
\begin{equation}
\begin{array}{l}
\bm D +\bm v \times\bm H/{c^2} =\bm \varepsilon \left( {\bm E +\bm v \times\bm B} \right),\\
\bm B +\bm E \times\bm v/{c^2} =\bm \mu \left( {\bm H +\bm D \times\bm v} \right),
\end{array} 
\end{equation}
where $\bm E$, $\bm D$, $\bm B$, and $\bm H$ are electromagnetic fields, $c$ is the speed of light in vacuum. Equation (1) amounts to introducing  bianisotropic response to the rotating cylinder, and the general form of the effective constitutive relation is given by
$\left( {\begin{array}{*{18}{c}}
\bm D\\
\bm B
\end{array}} \right) = \left( {\begin{array}{*{18}{c}}
{\bar{\bm\varepsilon} }&{{{\bar{\bm \chi} }_{eh}}}\\
{{{\bar{\bm \chi} }_{he}}}&{\bar {\bm\mu} }
\end{array}} \right) \left( {\begin{array}{*{18}{c}}
\bm E\\
\bm H
\end{array}} \right)$, where $\bar {\bm\varepsilon}  = {\varepsilon _0}{{\bar {\bm\varepsilon} }_r}$,$\bar {\bm\mu}  = {\mu _0}{{\bar {\bm\mu} }_r}$,${{\bar {\bm\chi} }_{eh}} = \sqrt {{\varepsilon _0}{\mu _0}} \bar {\bm\chi} _{eh}^r$ and ${{\bar {\bm\chi} }_{he}} = \sqrt {{\varepsilon _0}{\mu _0}} \bar {\bm\chi} _{he}^r$. Notably, all the elements of the material tensor explicitly depend on rotation speed $\Omega$. In contrast to reciprocal medium which have $\bar {\bm\varepsilon}  = {{\bar {\bm\varepsilon} }^T}, \bar {\bm\mu}  = {{\bar {\bm\mu} }^T}$, and ${{\bar {\bm\chi} }_{eh}} =  - \bar {\bm\chi} _{he}^T$, the rotating cylinder does not obey reciprocity, and the bianisotropic tensor elements satisfy  $\bar {\bm\varepsilon}  = {{\bar {\bm\varepsilon} }^T},\bar {\bm\mu}  = {{\bar {\bm\mu} }^T},{{\bar {\bm\chi} }_{eh}} =  \bar {\bm\chi} _{he}^T$. The rotation introduces non-reciprocity to the equivalent electromagnetic medium, therefore, it can be employed to realize  the optical isolation\cite{maayani2018flying,huang2018nonreciprocal}.

The source-free Maxwell's equations with time harmonic dependence ${e^{-i\omega t}}$ are given by $\left( {\bm\nabla  \times  - i{k_0}{{\bar {\bm\chi} }_{he}^r}} \right)\bm e = i{k_0}{\bar {\bm\mu}_r }\bm h$ and $\left( {\bm\nabla  \times  + i{k_0}{{\bar {\bm\chi} }_{eh}^r}} \right)\bm h =   -i{k_0}{\bar {\bm\varepsilon}_r }\bm e $,  where $k_0=\omega\sqrt{\varepsilon_0 \mu_0}$, $\omega$ is the angular frequency, $\bm e=\bm E,\bm h =Z_0 \bm H$ are the normalized electromagnetic field, and $Z_0$ is the vacuum  impedance. Considering TE modes where the electric field only has  $e_z$ component in a 2D cylindrical rotating cavity, see Fig. 1(a), one arrives at the Helmholtz equation by eliminating the magnetic field $\bm h$ as follows,
\begin{equation}\label{Helmholtz_equation}
\left( {\frac{1}{r}\frac{\partial }{{\partial \theta }} + i{k_0}A_{\theta}} \right)^2 {e_z}  +\frac{\mu_{rr}}{{{\mu _{\theta \theta }}}} \frac{1}{r}\frac{\partial }{{\partial r}}\left( {\frac{\partial }{{\partial r}}{e_z}} \right) + k_0^2{\varepsilon _{zz}}{\mu_{rr}}{e_z} = 0,
\end{equation}
where $ A_{\theta} = -\frac{{cr\Omega( \varepsilon_r \mu_r-1 ) }}{{ c^2 - \varepsilon_r \mu_r  {r^2}{\Omega ^2}}}$,
${\mu _{rr}}=\frac{\mu_r\left(c^2-r^2\Omega^2\right)}{c^2-\varepsilon_r\mu_r r^2\Omega^2}$, ${\mu _{\theta\theta }}=\mu_r$, and ${\varepsilon _{zz}}=\frac{\varepsilon_r\left(c^2-r^2\Omega^2\right)}{c^2-\varepsilon_r\mu_r r^2\Omega^2}$. Importantly, \Eq{Helmholtz_equation} gives the explicit expression of  the gauge field (i.e. vector potential) in the azimuthal direction: $\bm A = A_{\theta}\hat {\bm e}_\theta$, which accounts for the rotation-induced effects. The effective magnetic field $ \mathcal{B}$ along $z$ axis can be obtained from the vector potential as
\begin{equation}\label{magnaticField}
\mathcal{B} = \nabla  \times\bm A = \frac{{2\Omega\left({\varepsilon _r}{\mu _r} - 1\right) }}{{{{c\left(1- {{\varepsilon _r}{\mu _r}{r^2}{\Omega ^2}/{c^2}} \right)}^2}}}\hat {\bm e}_z=B_z\hat {\bm e}_z,
\end{equation}
where $\hat {\bm e}_z$ is the unit vector in $z$ direction. 

\subsection{Sagnac frequency splitting}
A cylindrical cavity of radius $R$ has a pair of CW and CCW modes with the mode electric field  ${e_z} = {E_z}\left( r \right){e^{ \pm i{k_\theta}r\theta }}$, where $k_\theta=m/r$, $m$ is an integer. Substituting it into \Eq{Helmholtz_equation}, one obtains the frequency splitting between CW and CCW modes as
\begin{equation}\label{deltaomega1}
\Delta \omega  =  \frac{{2{k_\theta }A_{\theta}c}}{{\varepsilon _{zz}\mu_{rr} - {A_{\theta}}^2}}.
\end{equation}
In the limit of $R\Omega \ll c$, all high-order terms about $\Omega$ in \Eq{deltaomega1} can be omitted, so $\varepsilon_{zz}\mu_{zz}-A_{\theta}^2\approx \varepsilon_r\mu_r$, the vector potential is $A_{\theta}\approx \tilde{A}_{\theta}= \frac{R\Omega\left(\varepsilon_r\mu_r-1\right)}{c}$, and effective magnetic field is $B_z \approx \tilde{B}_z=\frac{\Omega\left(\varepsilon_r\mu_r-1\right)}{c}$. We then get
\begin{equation}\label{deltaomega2}
\Delta \omega =\frac{2k_{\theta}\tilde{A}_{\theta}c}{\varepsilon _r\mu_r} =\frac{2m\tilde{B}_{z}c}{\varepsilon _r\mu_r}=\frac{{2m({\varepsilon _r\mu_r} - 1)\Omega }}{{{\varepsilon _r\mu_r}}},
\end{equation}
which describes the Sagnac effect induced by the effective magnetic field in a closed cavity. From \Eq{deltaomega2}, it's clear that the frequency splitting of a pair of CW and CCW modes is proportional to the azimuthal quantum number $m$ and angular velocity $\Omega$.  For comparison, a set of representative researches on Sagnac frequency splitting are listed in Table.1 in Appendix A, including inertial frame and co-rotating frame. Surprisingly, \Eq{deltaomega2} shows good consistency with the formula in second row of Table.1 if the dispersion term is omitted. The explicit expression of rotation-induced effective magnetic field in \Eq{magnaticField} and the rigorous  relation  between the Sagnac frequency splitting and the rotation-induced effective gauge field in \Eq{deltaomega1}  are the first main result of our  paper.

\begin{figure}[t]\label{fig1}
\centering
\includegraphics[width=13.3cm,height=6.6cm]{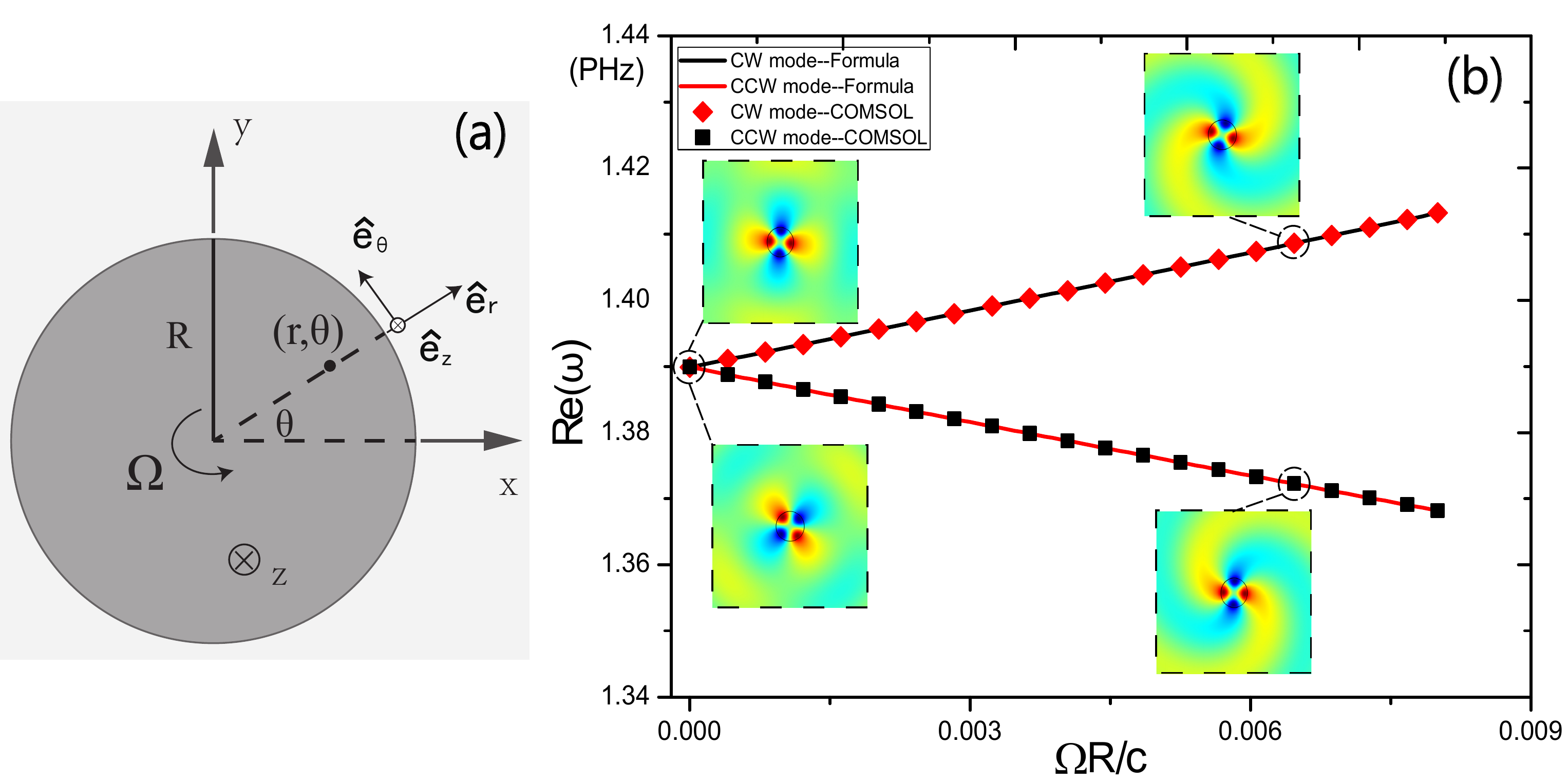}
\caption{(a) The 2D structure of a uniformly rotating
dielectric cylinder with angular speed $\Omega$. (b) Comparison of results on Sagnac frequency splitting obtained by COMSOL simulation and the analytical expression of \Eq{deltaomega1}. The insets show the $E_z$ components of the modal profiles of CW and CCW mode at $\Omega R/c=0$ and $\Omega R/c=0.00646$ obtained by COMSOL simulation. For the simulation, we use vacuum wavelength $\lambda_0=1550nm$, cylinder radius $R=200nm$ and the relative permittivity $\varepsilon_r=16$ for the stationary cavity. }
\end{figure}

\section{Results and discussions}
\subsection{Frequency splitting simulation} 
We examine the  Sagnac frequency splitting of CW and CCW modes using the analytical results of \Eq{deltaomega1} and the full-wave simulation results obtained from finite-element calculation. Though the rotating cylindrical rod can be calculated analytically\cite{movassagh2013optical,cao2015dielectric,ge2015rotation,sunada2007design}, the finite-element model we adopted here can handle generically bianisotropic medium with complex geometric shape\cite{dong2017valley}. COMSOL Multiphysics with the modified weak form of finite-element formulation\cite{knuthwebsite,sakoda2004optical,fushchich2013symmetries,xu2015elimination} is used to simulate the rotating cylinder in this paper. Figure 1(b) shows the real parts of eigenfrequencies of CW and CCW modes with  azimuthal number $m=2$  as a function of dimensionless angular speed $\Omega R/c$, where the solid lines denote the analytical results from \Eq{deltaomega1} and the symbols denote the simulation results. The analytical results match perfectly well with  the simulation results, indicating a linear dependence as $\Omega R/c$. The insets in Fig. 1(b) show the modal  profiles of the CW and CCW modes. We note that the mode  pair (i.e. CW and CCW modes)  are originally degenerated at $\Omega=0$. As $\Omega R/c$ increases, this degeneracy is lifted, and the profiles of CW and CCW mode carry a 'spinning' chiral feature, as shown by the right insets in Fig. 1(b). It is worthy to point out that the cavity modes examined in this paper is essentially quasi-normal modes with complex eigenfrequencies and finite Q-factors. The finite value of the Q-factor approximately determines the line-shape of cavity resonance, and it turns out to be rather important in determining the modal hybridization and thus the far-field pattern at lower rotation speed, which will be discussed shortly in the next section.

\subsection{Hybridization of CW and CCW components in a rotating cylinder} 

Sagnac frequency shift reflects the modification of the eigenfrequency induced by the rotation.  Now we proceed to discuss the impact of rotation upon the eigenfields. It is well understood that the eigenfields for the two split bands due to rotation, as shown in Fig. 1(b), are CW  and CCW modes respectively, i.e., $\left| {CW} \right\rangle = {E_z}\left( r \right){e^{im\theta }}$ and $\left| {CCW} \right\rangle = {E_z}\left( r\right){e^{-im\theta }}$. However, a precise one-to-one correspondence between the upper/lower band to the CW/CCW mode only exists at high rotation speed. As discussed in details by Ge\cite{ge2015rotation}, a low rotation speed gives rise to a rotation-dependent far-field distribution, indicating a certain mixture of CW and CCW mode.

\begin{figure}[t]\label{fig2}
\centering
\includegraphics[width=12cm,height=12cm]{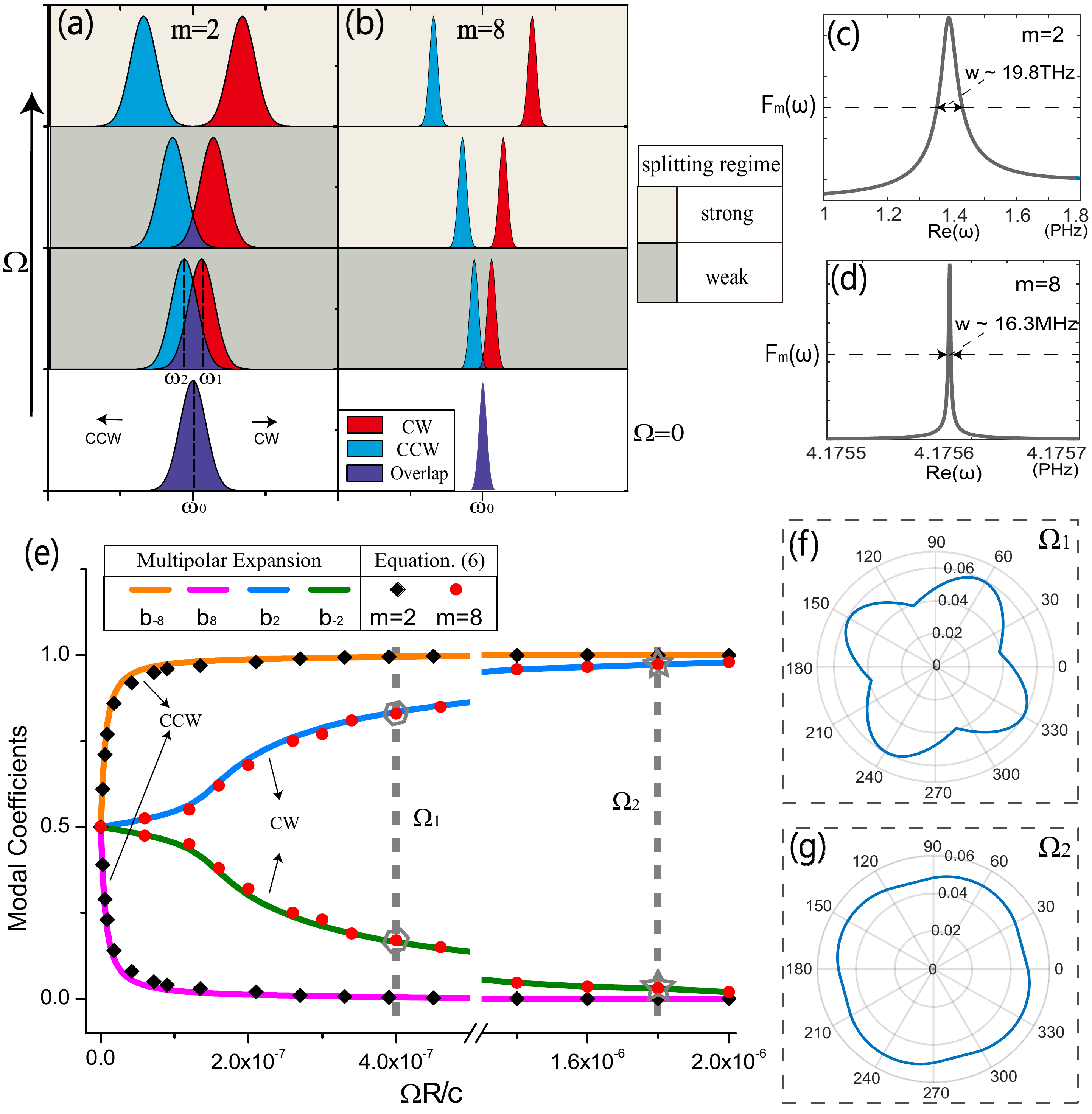}
\caption{(a)-(b) Schematic diagrams of the effect of Sagnac frequency splitting on CW/CCW modes for a rotating cavity. Line-shape functions of the modes at $\Omega=0$ with azimuthal number (c) $m=2$ and (d) $m=8$ respectively. (e) The variation of modal coefficients obtained by the multipolar expansion and Eq. (6) for CW mode with azimuthal number $m=2$ and CCW mode with $m=8$ versus the dimensionless angular speed $\Omega R/c$. The setup of parameters is shown in Fig. 1. Far-field distribution obtained by superposing the eigenfields $\left| {CW} \right\rangle$ and $\left| {CCW} \right\rangle$ with corresponding ratio at (f) $\Omega_1 R/c=
4 \times {10^{ - 7}}$ and (g) $\Omega_2 R/c=
1.8 \times {10^{ - 6}}$.}
\end{figure}

To reveal the underlying principle behind rotation-dependent far-field distribution, we first identity the modal splitting of $\left| {CW} \right\rangle$ and $\left| {CCW} \right\rangle$ as the rotation speed $\Omega$ varies. At $\Omega=0$, as illustrated in Figs. 2(a)-2(b), each resonance of the stationary rod has a finite broadening, and the y dimensions represent the normalized resonant amplitude. The full width at half maximum (FWHM), denoted as $w$ in this paper, decreases as the azimuthal quantum number $m$ increases. In the presence of non-zero rotation speed, i.e., $\Omega > 0$, the coupling between the original eigenmodes $\left| {CW} \right\rangle$ and $\left| {CCW} \right\rangle$ does not vanish, and there is modal splitting $\Delta \omega$ at this situation, as shown in Figs. 2(a)-2(b). Depending  on the relative values between $w$ and $\Delta \omega/2$, the modal splitting can be categorized into two regimes, i.e., weak splitting  regime and strong splitting regime, as shown in Figs. 2(a) and 2(b).  In strong splitting  regime, i.e., $\Delta \omega/2 > w$, the two resonant modes are completely split with negligible spectra overlapping. In contrast, in the weak splitting regime, there is  spectral overlapping due to the fact that $\Delta \omega/2 < w$, which gives rise to the rotation-dependent far-field distribution as discussed by Ge\cite{ge2015rotation}. The above analysis gives a qualitative understanding of how $\left| {CW} \right\rangle$ and $\left| {CCW} \right\rangle$ contribute to the eigenfields of the rotating cavity. In the following, we give a quantitatively assessment of   how $\left| {CW} \right\rangle$ and $\left| {CCW} \right\rangle$ contribute to the eigenfields of a 2D rotating cylinder by including the line-shape function explicitly in the weak splitting regime.

Due to the cylindrical symmetry, the eigenfield  $\left| \varphi  \right\rangle$ of the stationary rod at $\Omega$=0 can be expressed as $\left| \varphi  \right\rangle  = f_{+m}( \omega ,\omega _0)\left| {CW} \right\rangle  + f_{-m}( \omega ,\omega _0 )\left| {CCW} \right\rangle$, where $f_{+m}( \omega ,\omega _0)$ ($f_{-m}( \omega ,\omega _0 )$)  is the line-shape function of CW (CCW) mode with resonant frequency $\omega_0$ and FWHM $w$ (see the definition and detailed calculation of  $f_{\pm }( \omega ,\omega _0)$ in  Appendix B). At $\Omega=0$, it can be proved that  $f_{+m}( \omega ,\omega _0)$ is identical to  $f_{-m}( \omega ,\omega _0 )$, i.e., $F_{m}( \omega ,\omega _0)=f_{+m}( \omega ,\omega _0)=f_{-m}( \omega ,\omega _0)$, so $F_m$ is line-shape function of the stationary cavity mode with azimuthal number $m$. The precisely calculated line-shape functions of the modes with $m=2$ and $m=8$ for the stationary rod are shown in Fig. 2(c) and 2(d), respectively, where the FWHM $w$ is on the order of  THz/MHz range. At low rotation speed ($\Delta \omega/2 < w$),  to a good approximation, the weighting factors of  $f_{\pm }( \omega ,\omega _0)$ and the line-shape function $f_{\pm }( \omega ,\omega _0)$ remain unchanged, except that the resonant frequencies are shifted by  half of the Sagnac frequency splitting. In this regard, the eigenfield of the low-speed rotating cylinder can be well approximated as follows, 
\beq\label{quasimode}
\left| \varphi  \right\rangle  = F_{m}( \omega ,\omega _0 +{\Delta\omega}/2 )\left| {CW} \right\rangle  + F_{m}( \omega ,\omega _0 -{\Delta\omega}/2)\left| {CCW} \right\rangle,
\eeq
where $\Delta \omega$ is the Sagnac frequency shift defined in Eq. (4). Evidently, the modal coefficients of the CW/CCW resonance at a certain eigenfrequency $\omega$ can be semi-analytically determined by \Eq{quasimode} in the weak splitting regime. This is the second main result of this paper.

In the following, we apply \Eq{quasimode} to quantitatively study the modal hybridization of the rotating cylinder, accompanied with the numerical verification from multipolar expansion analysis, as shown in Fig. 2(e). Based on the revised finite-element calculation as well as Mie scattering theory\cite{bohren2008absorption,kozaki1982scattering,grahn2012electromagnetic}, the multipolar components of the eigenfields can be extracted numerically by projecting the eigenfields of the rotating cylinder to vector spherical harmonic waves. Surprisingly, the modal coefficients obtained by the full-wave simulation and multipolar expansion agree with the results of our simple analytical model given by \Eq{quasimode}, for both the CW mode (orange and magenta  for $|m|$=8) and the CCW mode (blue and  olive for $|m|$=2), as shown in Fig. 2(e). At $\Omega$=0, it is easy to find that ${b_{ - m}} =  {b_m}$, which is consistent with \Eq{quasimode} at $\Delta \omega$=0. As the cylinder rotates, the coefficients of  the original equal pair of multipolar components for CW (CCW) mode start to differ, with $b$ coefficients increased for $m$>0 ($m$<0) and decreased for $m$<0 ($m$>0). In comparison, the variation trends of the multipolar components of the CW mode and that of the CCW mode are exactly the opposite. This can be understood by our hybridization model that the Sagnac frequency splitting gives rise to  a dominant role of  $\left| {CW} \right\rangle$($\left| {CCW} \right\rangle$) in CW (CCW) modes, such that the weighting factor of $\left| {CW} \right\rangle$($\left| {CCW} \right\rangle$) increases (decreases) with angular velocity in CW modes while decreases (increases) in CCW modes. Evidently, $b$ coefficients of the CW mode with $m<0$ and $b$ coefficients of the CCW mode with $m>0$ drop to 0 at high rotation speed due to the fact that the Sagnac frequency shift is so large that there is negligible spectra overlapping. We also note that the multipolar components of the cavity mode with large $m$ value  are more sensitive to the rotation speed $\Omega$ than that of the cavity mode with small $m$ value. This is due to a simple fact that the line-shape function of a higher-order mode has a smaller $w$, and $\Delta \omega$ with same $\Omega $ can easily lead to complete splitting of the higher-order CW and CCW modes. 

According to our analysis, the rotation-dependent far-field distribution can be obtained by superposing the eigenfields $\left| {CW} \right\rangle$ and $\left| {CCW} \right\rangle$ with proper weighting factors in accordance with the angular velocity $\Omega$.  Taking the CW mode with $m=2$ in Fig. 2(e) as an example, the far-field distribution at $\Omega_1 R/c =4 \times {10^{ - 7}}$ and $\Omega_2 R/c =1.8 \times {10^{ - 6}}$ are shown in Fig. 2(f) and 2(g) respectively.  From Fig. 2(f) to Fig. 2(g), the original four-lobed shape gradually becomes a circle, which is consistent with our previous analysis.  As long as the angular speed $\Omega$ is large enough, the far field has a circular pattern, indicating that only $\left| {CW} \right\rangle$  component exists.  

\section{Conclusion}

In conclusion, we have investigated the optical properties of a rotating cavity beyond the first order approximation. Starting from the first principle of electromagnetism, we have derived the effective magnetic field and Sagnac frequency splitting in a 2D rotating rod from the perspective of synthetic gauge field. With the modified weak form of finite-element model, full-wave simulations are implemented to examine the  rotating cavity by treating it as a stationary bianisotropic medium. Moreover, a simple hybridization model is semi-analytically constructed for the cylinder at a low rotation speed, which provides a simple yet accurate description of the rotation-dependent far-field pattern. Our model is benchmarked  against numerical analysis of multipolar expansions and enables a quantitative study of the far-field distribution of the cylinder at different angular velocity $\Omega$.
The results might be applied in the angular speed measurement of rotating devices by monitoring the modification of the far-field distribution.

\section*{Appendix A: Summary of different formulae of Sagnac frequency shift}
\begin{table}[htbp!]
\centering
\caption{Various expressions for Sagnac frequency splitting of optical rotating cavities}
\resizebox{\textwidth}{1.8cm}{
\begin{tabular}{|c|c|c|}  
\hline
Formula & Approximation Method & Reference Frame \\  
\hline       
$\Delta\omega=\frac{2m\Omega}{n^2}$\cite{sunada2007design,scheuer2007direct,sarma2012wavelength,cao2015dielectric}$\Leftrightarrow$$\Delta\nu=\frac{2R\Omega}{n{\lambda}_0}$\cite{post1967sagnac,malykin2014sagnac,liang2017sagnac}  & the wave equation& corotating RF \\
\hline
$\Delta\omega=2\omega_{rest}\frac{nR\Omega}{c}(1-\frac{1}{n^2}-\frac{\lambda}{n}\frac{dn}{d\lambda})$\cite{malykin2000sagnac,jing2018nanoparticle,maayani2018flying} & geometrical optics  & inertial RF \\
\hline
$\Delta\nu=\frac{2R\Omega}{\lambda_0}$\cite{rosenthal1962regenerative,meyer1983passive,mignot2009single}  & experiment & inertial RF \\
\hline
$\Delta\nu=\frac{2nR\Omega}{\lambda_0}$\cite{anderson1969electromagnetic}  & the wave equation  & {static medium rotating RF} \\
\hline
$\Delta\nu=\frac{\Omega D}{c}\nu_{rest}$\cite{heer1964resonant,cheo1964beat} &Calibration wave function  & inertial RF\\
\hline
\end{tabular}
\label{difference}
}
\end{table}

\section*{Appendix B: Line-shape function of stationary cylindrical cavity modes}
Starting from the Mie scattering theory, when the incident light is normal to the cylinder axis, the coefficient of the scattered field $a_m$ vanishes, and $b_m$ has the following form
\begin{equation}
{b_m} = \frac{{{J_m}\left( {n{\rho _0}} \right){J_m'}\left( {\rho _0} \right) - n{J_m'}\left( {n{\rho _0}} \right){J_m}\left( {\rho _0} \right)}}{{{J_m}\left( {n{\rho _0}} \right)H_m^{\left( 1 \right)'}\left( {\rho _0} \right) - n{J_m'}\left( {n{\rho _0}} \right)H_m^{\left( 1 \right)}\left( {\rho _0} \right)}},
\end{equation}
where $J_m$ and $H_m^{\left(1\right)}$ are Bessel function of first kind and Hankel function, respectively, of integral order $m$ (i.e. angular quantum number). $n$ is the relative refractive index of the cylinder. ${\rho _0}={k_0}R$, $k_0=\omega/c$ is the wave number in vacuo,
$\omega$ being the frequency in the time harmonic oscillating field ${e^{i\omega t}}$, and $c$ is the speed of light in vacuum. $R$ represents the radius of the cylinder. We define the line-shape function as $f(\omega)=1/\left({{{J_m}\left( {n{\rho _0}} \right)H_m^{\left( 1 \right)'}\left( {\rho _0} \right) - n{J_m'}\left( {n{\rho _0}} \right)H_m^{\left( 1 \right)}\left( {\rho _0} \right)}}\right)$, using the identities $J_m'{\left( x \right)} = \frac{1}{2}\left[ {{J_{m - 1}}\left( x \right) - {J_{m + 1}}\left( x \right)} \right]$ and 
$H_m^{\left( 1 \right)'}\left( x \right) = \frac{1}{2}\left[ {H_{m - 1}^{\left( 1 \right)}\left( x \right) - H_{m + 1}^{\left( 1 \right)}\left( x \right)} \right]$ , by varying angular frequency $\omega$, the absolute value of the $f(\omega)$ function of different angular quantum number $m$ are obtained. 

As a test, we make a comparison with the results of COMSOL. For example, for the mode with angular quantum number $m=2$ at $\Omega=0$, its (angular) eigenfrequency is
$1.389925 \times {10^{15}} + 1.982725 \times {10^{13}}i$, for the other mode with angular quantum number $m=8$ at rest, its eigenfrequency is
$4.175587 \times {10^{15}} + 1.627345 \times {10^{7}}i$. However, for the CCW mode with azimuthal number $m=2$ at $\Omega R/c=1 \times {10^{ - 6}}$, its eigenfrequency becomes 
$1.389923 \times {10^{15}} + 1.982726 \times {10^{13}}i$.
By approximation,  we assume that the imaginary parts of the modes remain unchanged, since the value of $\Omega R/c$ is particularly small($ \sim  {10^{ - 6}}$) in our considered cases and the real parts of the modes are at least 70 times that of  imaginary parts.

\section*{Funding}
Natural National Science Foundation (NSFC) (11874026, 61735006, 61775063); National Key Research and Development Program of China (2017YFA0305200); Research Grants Council of Hong Kong SAR (CityU 21302018).

\section*{Acknowledgment}
We thank Prof. Wei Liu for fruitful discussions.


\addcontentsline{toc}{chapter}{Bibliography}
\bibliography{sample}


\end{document}